\documentclass[iop]{emulateapj}
\usepackage{apjfonts}
\def\Sref#1{Sect.~\ref{sec:#1}}
\def\Fref#1{Figure~\ref{fig:#1}}

\slugcomment{Revised version, accepted for publication in
  Astrophysical Journal (Letters), 10 September 2010}
\begin{document}
\title{Sub-Chandrasekhar White Dwarf Mergers as the Progenitors of
 Type Ia Supernovae}
\author{Marten H. van Kerkwijk\altaffilmark{1,2,3}, 
        Philip Chang\altaffilmark{4},
    and Stephen Justham\altaffilmark{2}}
\altaffiltext{1}{Department of Astronomy and Astrophysics, University
  of Toronto, 50 St. George Street, Toronto, ON M5S 3H4, Canada; 
  \email{mhvk@astro.utoronto.ca}}
\altaffiltext{2}{Kavli Institute for Astronomy and Astrophysics,
  Peking University, Beijing, China} 
\altaffiltext{3}{Caltech Optical Observatories 249--17, California
  Institute of Technology, Pasadena, CA 91125}
\altaffiltext{4}{Canadian Institute for Theoretical Astrophysics, 60
  St George Street, Toronto, ON M5S 3H8, Canada}

\begin{abstract}
  Type Ia supernovae are generally thought to be due to the
  thermonuclear explosions of carbon-oxygen white dwarfs with masses
  near the Chandrasekhar mass.  This scenario, however, has two
  long-standing problems.  First, the explosions do not naturally
  produce the correct mix of elements, but have to be finely tuned to
  proceed from sub-sonic deflagration to super-sonic detonation.
  Second, population models and observations give formation rates of
  near-Chandrasekhar white dwarfs that are far too small.  Here, we
  suggest that type Ia supernovae instead result from mergers of
  roughly equal-mass carbon-oxygen white dwarfs, including those that
  produce sub-Chandrasekhar mass remnants.  Numerical studies of such
  mergers have shown that the remnants consist of rapidly rotating
  cores that contain most of the mass and are hottest in the center,
  surrounded by dense, small disks.  We argue that the disks accrete
  quickly, and that the resulting compressional heating likely leads
  to central carbon ignition.  This ignition occurs at densities for
  which pure detonations lead to events similar to type Ia supernovae.
  With this merger scenario, we can understand the type Ia rates, and
  have plausible reasons for the observed range in luminosity and for
  the bias of more luminous supernovae towards younger populations.
  We speculate that explosions of white dwarfs slowly brought to the
  Chandrasekhar limit---which should also occur---are responsible for
  some of the ``atypical'' type Ia supernovae.
\end{abstract}


\section{Introduction}

Type Ia supernovae (SN~Ia) result from thermo-nuclear explosions of
carbon-oxygen white dwarfs (CO~WDs).  They are generally thought to be
triggered when the WD approaches (for accretion) or exceeds (for a
merger) the Chandrasekhar mass, and the density and temperature become
high enough to start runaway carbon fusion.  This scenario, however,
neither naturally leads to explosions that reproduce the observed
lightcurves and remnants, nor easily accounts for the variation in
SN~Ia properties and their dependence on host galaxy.  Furthermore,
the predicted formation rates are lower than observed.

The above leads us to reconsider the assumptions underlying the
standard picture.  After reviewing the salient properties of SN~Ia
(\Sref{properties}), we first argue that their rates are easiest to
understand if most mergers of CO~WDs lead to SN~Ia, independent of
whether or not the total mass exceeds the Chandrasekhar mass.  After a
brief discussion of previous sub-Chandrasekhar models
(\Sref{explosions}), we next argue that ignition following mergers is
likely (\Sref{ignition}).  We close with some ramifications
(\Sref{conclusions}).

\section{Properties of individual SN~Ia and their remnants}
\label{sec:properties}

The lightcurves of most SN~Ia are remarkably similar, and can
be described empirically as a (nearly) single-parameter family, in
which timescale and maximum luminosity are tightly correlated, with
longer-lasting explosions being more luminous and energetic
\citep{phil93}.  Underlying this variation is the amount of
radioactive $^{56}$Ni.  From SN~Ia spectra, \cite{mazz+07} find that
this ranges from $\sim\!0.1$ to $0.9\,M_\odot$.  They also infer
$\sim\!0.1\,M_\odot$ of stable iron-peak elements, and an amount of
intermediate-mass elements that is such that the total mass of nuclear
processed material is roughly constant, just over $1\,M_\odot$.
\cite{stri+06} use peak luminosities to infer similar $^{56}$Ni
masses, but their total masses, inferred from the times that the
ejecta become optically thin, do not cluster, but range from 0.5 to
$1.3\,M_\odot$.

From SN~Ia spectra, it is also clear that the ejecta are stratified,
with iron-peak elements formed deeper inside \citep{mazz+07}, and
(small amounts of) unprocessed carbon on the outside \citep{thom+07}.
Hydrogen is absent \citep{leon07}, inconsistent with expectations for
a hydrogen-rich progenitor companion with a strong wind or easily
entrained envelope.

Studies of SN~Ia remnants paint a similar picture.
% (for a review, \citealt{bade10}).  
For instance, \cite{bade+06} find that for Tycho, models with about
twice as much iron-peak as intermediate-mass elements best reproduce
the X-ray spectrum, consistent with SN~1572A having been a
``standard'' SN~Ia (confirmed beautifully using light echoes;
\citealt{krau+08}).  From the ionization structure, \cite{bade+06}
infer stratified ejecta, strongly suggesting the explosion was
(partly) supersonic.  Comparing the predicted remnant flux,
Chandrasekhar-mass models are too luminous.  \cite{bade+06} attribute
this to a break-down in their one-dimensional remnant models, but it
could also indicate a lower mass.  A separate clue is that most
remnants appear to be evolving into a constant-density medium, with
properties like those of the warm interstellar phase, and unlike those
expected for progenitors with strong, fast winds \citep{bade+07}.

The rates and properties of SN~Ia depend on environment, with
star-forming galaxies having higher rates of, on average, brighter
SN~Ia than passively evolving galaxies (e.g.,
\citealt{hamu+95,mann+05,sull+10}).  The rate has been suggested to
depend on both mass and star-formation rate \citep{mann+05},
or to simply be a roughly constant fraction, of $\sim\!1$\%, of the
instantaneous WD formation rate (\citealt{priths08}; but see
\citealt{maoz+10}, whose data suggest a break beyond $\sim\!2\,$Gyr
[their Fig.~5]).  Consistent with the latter, \citet{rask+09} found
that SN~Ia are delayed by $\sim\!200$ to 500\,Myr from the onset of
star formation (as inferred from local environments).  Integrated over
a Hubble time, $\sim\!0.0023\pm0.0006$ SN~Ia seem to occur for every
solar mass formed \citep{mann+05,maoz+10}, with even higher numbers,
of $\gtrsim\!0.0034$ inferred from galaxy cluster iron abundances
\citep{maozsg10}.

In summary, the lightcurves, spectra, and remnants of SN~Ia supernovae
seem to require center-lit explosions of CO~WDs, with masses of
$\gtrsim\!1\,M_\odot$.  Their progenitor systems likely did not host
companions with strong hydrogen-rich winds or loosely bound envelopes.
The rate of SN~Ia appears to be $\sim\!1$\% of the WD formation rate,
and younger systems produce more luminous explosions.

\section{Expected SN~Ia rates}
\label{sec:rates}

The number of SN~Ia per solar mass formed, $\sim\!0.0023$, is higher
than expected for Chandrasekhar-mass systems, both from counts of
suitable intermediate-mass progenitors (e.g., \citealt{maoz08}) and
from population synthesis calculations (for recent work, see, e.g.,
\citealt{ruitbf09,menn+10}).  The details are complex and metallicity
dependent, but below we elucidate the issues with rates estimated from
basic principles (for a more formal analysis, see \citealt{greg05}).

For the primary in an interacting binary to leave a CO~WD, it must be
massive enough not to leave a He~WD, but not so massive that it forms
an ONe~WD or neutron star.  Using Figs.~1 and~2 of \cite{webb08}, we
estimate a mass range $1.8\lesssim{}M_1\lesssim7\,M_\odot$ (the lower
limit exceeds that for single stars because the star needs to ignite
helium after its interaction).  A \cite{chab05} initial mass function
produces $n_{1.8<M_1<7}=0.067$ such stars per solar mass formed.  To
produce a CO~WD of $\gtrsim\!0.7\,M_\odot$ (half the Chandrasekhar
mass), a $\gtrsim\!3.5\,M_\odot$ primary is required, of which only
$n_{3.5<M_1<7}=0.020$ are formed.  For these suitably massive stars, a
fraction $f_{\rm{}bin}\simeq\frac{2}{3}$ is in binaries, and, of those, a
fraction $f_{10<P<2000}\simeq0.2$ have periods between 10 and 2000\,d
\citep{duqum91}, such that they interact ($P\lesssim2000\,$d), but
only after the main sequence, with a fully formed helium core
($P\gtrsim10\,$d).

The further evolution depends on whether mass transfer is stable or
not.  If it is unstable, common-envelope evolution will drastically
shrink the orbit, leading to possible further evolution via the
single-degenerate channel \citep{wheli73}.  If it is stable, the
system remains wide and further evolution leads to a second CO~WD as
well as a second, unstable mass-transfer phase that shrinks the orbit,
as required for the double-degenerate scenario \citep{webb84,ibent84}.

For the giants considered here, mass transfer to a less massive
companion is generally expected to be dynamically unstable (e.g.,
\citealt{webb08}).  So one na\"{i}vely expects a small fraction
$f_{\rm{}wide}$ left in wide orbits, and a near-unity fraction
$f_{\rm{}close}$ left in close orbits.  Empirically, however, the
first mass-transfer phase sometimes leaves wide orbits
(\citealt{nele+00}; for a discussion, see \citealt{webb08}),
presumably for nearly equal-mass binaries (which may be relatively
common; \citealt{pinss06}).  Indeed, the existence of fair numbers of
both double-degenerate binaries and cataclysmic variables suggests
neither fraction is small.  Below, we assume
$f_{\rm{}wide}\simeq\frac{1}{3}$ and
$f_{\rm{}close}\simeq\frac{2}{3}$; likely, neither is off by more
than~50\%.

\subsection{Single degenerates}
\label{sec:sd}

In principle, the number of CO WDs formed in close orbits with
non-degenerate companions,
$n_{1.8<M_1<7}f_{\rm{}bin}f_{10<P<2000}f_{\rm{}close}\simeq0.006$ per
solar mass, could reproduce the SN~Ia rate, and many routes to
explosions have been proposed \citep{ibent84}.  No route, however,
seems both common and efficient in growing the WD to the Chandrasekhar
mass.  If mass transfer is too slow, novae occur, which appear to
remove as much mass as was accreted (\citealt{townb04}; possible
counterexamples are RS~Oph and U~Sco).  If it is faster, hydrogen
burns stably, but only a small range avoids expansion and mass loss
\citep{nomo+07}.

Empirically, the only efficient systems appear to be the supersoft
sources \citep{rappdss94}, but those are far too rare
\citep{dist10,gilfb10}.  We may be missing systems, e.g., rapidly
accreting WDs that expanded and hid from X-ray view.  However, for
such sources---as for many single-degenerate channels---the absence of
evidence for hydrogen and wind-blown bubbles is surprising.  The lack
of convincing solutions to these issues motivates us to look for
alternative progenitors.

\subsection{Double degenerates}
\label{sec:dd}

Given the near-unity mass ratio required to keep a wide orbit in the
first mass-transfer phase, the two CO~WDs are expected to have similar
masses.  To estimate the number of double degenerates with total mass
exceeding the Chandrasekhar mass, we thus use the number of binaries
with sufficiently massive primaries,
$n_{3.5<M_1<7}f_{\rm{}bin}f_{10<P<2000}f_{\rm{}wide}\simeq0.0009$ per
solar mass formed.  This is less than half the required number, which
poses a significant problem, especially as some systems will be too
wide to merge in a Hubble time. 

Indeed, this realisation motivates our consideration of
sub-Chandrasekhar merger models: if \emph{all} mergers of CO~WDs would
lead to SN~Ia, one has
$n_{1.8<M_1<7}f_{\rm{}bin}f_{10<P<2000}f_{\rm{}wide}\simeq0.003$
possible progenitors, which is consistent with the observations.

Furthermore, sub-Chandrasekhar mergers could explain the observed
delay-time distribution.  Generally, distributions of formation times
are shallower, $\propto{}t^{-0.5}$ \citep{priths08}, than those of
merger times, $\propto{}t^{-1}$ (e.g., \citealt{greg05}).  Thus, one
expects the SN~Ia rate to scale with the WD formation rate.
Quantitatively, the scale factor is $\sim\!0.01$ \citep{priths08},
while the fraction of WDs formed in double degenerates is about
$f_{\rm{}bin}f_{10<P<2000}f_{\rm{}wide}\simeq0.04$.  Thus,
$\sim\!25$\% of the double degenerates should merge fast compared to
the progenitor lifetime.  But the scaling with WD formation rate will
hold only for a duration roughly equal to the lifetime of the
lowest-mass progenitor.  For $\gtrsim\!3.5\,M_\odot$ progenitors, this
will be $\sim\!200\,$Myr, much shorter than observed, but for
$\gtrsim\!1.8\,M_\odot$ stars, it is $\sim\!1.7\,$Gyr, consistent with
the observations.

\section{Previous Sub-Chandrasekhar Models}
\label{sec:explosions}

We are not the first to consider sub-Chandrasekhar models for SN~Ia.
\citet{woosw94} suggested that single-degenerate sub-Chandrasekhar
mass explosions might be triggered by detonations of overlying He
layers.  However, those lead to stratifications inconsistent with the
observations (see \Sref{properties}).

In contrast, \cite{sim+10} studied central detonations of
sub-Chandrasekhar WDs (ignoring the ignition mechanism, which we
address in \Sref{ignition}).  They found lightcurves and spectra
similar to SN~Ia, and showed that for more massive WDs, more iron-peak
elements were created, with
$M_{\rm{}Ni}\simeq\{0.06,0.32,0.58,0.85\}\,M_\odot$ for
$M_{\rm{}WD}=\{0.88,0.97,1.06,1.15\}\,M_\odot$.

That mass dependence arises because iron-peak elements are produced
only in regions that reach $\sim\!4\times10^9\,$K before degeneracy is
lifted, which requires a density $\rho\gtrsim10^7{\rm\,g\,cm^{-3}}$
(to produce intermediate mass elements requires
$\rho\gtrsim\!2\times10^6{\rm\,g\,cm^{-3}}$).  For the same reason,
detonations of near-Chandrasekhar WDs produce too much nickel: almost
the whole WD is above the critical density.  This conundrum can only
be solved by an initial deflagration phase, which allows the WD to
expand \citep{khok91}.  For lower mass WDs, this is not necessary.

\begin{figure}
\includegraphics[width=\hsize]{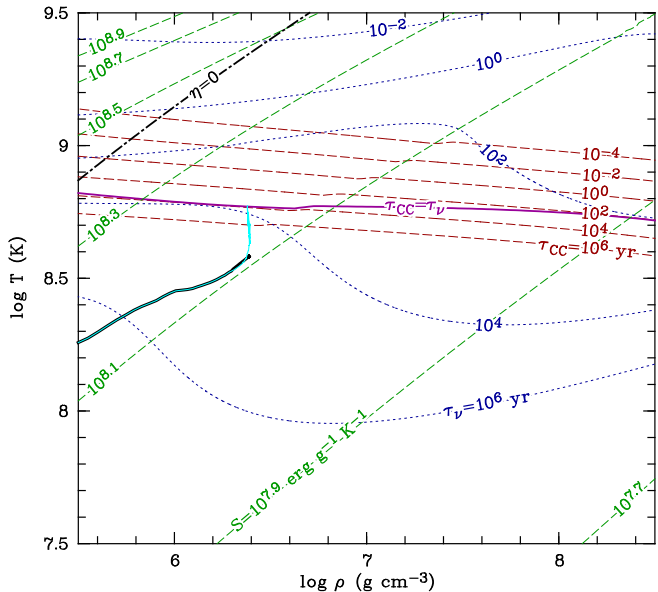}
\caption{Temperate-density profile of the remnant of a merger of two
  $0.6\,M_\odot$ CO~WDs.  The thin cyan curve shows the profile from
  the simulations of \cite{loreig09}, and the thick black curve the
  one that results from mixing the central convectively unstable
  region.  After its formation, the remnant will accrete and become
  denser and hotter, evolving at constant entropy (green dashed,
  diagonal contours) as long as the accretion timescale is shorter
  than $\tau_\nu=C_VT/\varepsilon_\nu$ and
  $\tau_{\rm{}CC}=C_VT/\varepsilon_{\rm{}CC}$, the timescales on which
  neutrinos can cool and carbon fusion can heat the core (blue dotted
  and red dashed contours, respectively).  The magenta solid curve
  shows where ignition occurs ($\tau_\nu=\tau_{\rm{}CC}$) and the
  thick black dot-dashed one where degeneracy is lifted (normalized
  electron chemical potential $\eta=0$).  (The entropies $S$, specific
  heats~$C_V$, and gain/loss rates~$\varepsilon$ were calculated using
  routines from the stellar evolution code {\sc mesa}
  [\citealt{paxt+10}; http://mesa.sourceforge.net];
  uncertainties in $\varepsilon_{\rm{}CC}$, while large, do not
  influence our conclusions)}
\label{fig:rhot}
\end{figure}

\section{Ignition}
\label{sec:ignition}

We argued that the SN~Ia rate and delay-time distribution could be
understood if mergers of CO WDs lead to SN~Ia even for
sub-Chandrasekhar total mass.  If mergers lead to explosions similar
to the detonations of sub-Chandrasekhar WDs, they will appear like
SN~Ia \citep{sim+10}.  Assuming more massive mergers also produce more
$^{56}$Ni, the range in luminosity and duration follows naturally, as
does the correlation with parent population age.

The critical remaining question is whether merger products become
sufficiently hot to ignite.  This has been addressed partly by recent
simulations, which include careful treatment of the equation of state
and of nuclear burning \citep{yoonpr07,loreig09,pakm+10}.  These show
qualitative differences between mergers of unequal and equal mass
binaries.\footnote{What constitutes ``equal'' is not yet known, though
  0.8 and $0.6\,M_\odot$ is not.  For reference, in the range
  0.6--0.9$\,M_\odot$, to have central densities within a factor of
  two requires $\Delta M\lesssim0.13\,M_\odot$.}  If one WD is
significantly lighter (and thus larger), mass transfer leads to its
total disruption, with the material wrapped around the more massive
companion; the merger product has a core that is cooler and rotates
more slowly than the envelope.  Any nuclear processing happened at the
core-envelope interface, as likely would any subsequent ignition
\citep{nomoi84,yoonpr07}.  Thus, we do not think such mergers lead to
SN~Ia.

If instead the two WDs have more equal masses, as expected for CO+CO
binaries (\Sref{dd}), the merger remnant is fully mixed and hottest in
the center.  Initially, it rotates differentially, but this is
dissipated, and one is left with a core holding $\sim\!80$\% of the
mass and rotating at a uniform rate near the mass-shedding limit,
surrounded by a somewhat sub-Keplerian, very dense, partially
degeneracy-pressure supported ``disk'' with a steep surface density
gradient ($\Sigma\propto{}r^{-5}$).

In the simulations of \cite{loreig09} and \cite{pakm+10}, the mergers
of equal-mass WDs do not become hot enough to ignite carbon burning,
except for masses above $\sim\!0.9\,M_\odot$.  The explosion of the
resulting $\gtrsim\!1.8\,M_\odot$ remnant, however, leads to a
subluminous SN~Ia.  This is not surprising: generally, merger remnants
have central densities similar to those of the (more massive of the)
pre-merger WDs.  For a $0.9\,M_\odot$ WD, the central density is
$\sim\!1.5\times10^7{\rm\,g\,cm^{-3}}$, and thus little $^{56}$Ni will
be produced (see \Sref{explosions}), leading to a subluminous
explosion.\footnote{The rare mergers of even more massive WDs should
  lead to more luminous explosions.}

We now turn to the simulation of the merger of two $0.6\,M_\odot$ WDs
of \cite{loreig09}.  This mass is close to the empirical mean mass of
WDs ($\sim\!0.65\,M_\odot$; \citealt{tremb09}), and thus, in our
picture, the merger should lead to a typical SN~Ia.  From the
simulation, the remnant's central density and temperature are
$\sim\!2.5\times10^6{\rm\,g\,cm^{-3}}$ and $\sim\!6\times10^8\,$K,
respectively.  This would appear to be close to what is required for
ignition, but since the central $\sim\!0.07\,M_\odot$ is convectively
unstable (\Fref{rhot}), the temperature will be reduced rapidly, on a
convective turn-over time.  The region's average entropy,
$s/k\simeq16$ per ion and $\sim\!1.3$ per electron, corresponds to a
core temperature of $\sim\!3.8\times10^8\,$K.

While this merger remnant is too cold (and insufficiently dense) to
produce a SN~Ia, it could explode later.  One possible heating
mechanism is the accretion of the thick disk, which will lead to an
increase in density and thus to compressional heating.  Assuming the
source remains at the mass-shedding limit and in roughly solid-body
rotation (e.g., due to magnetic fields; \Sref{complications} below),
the density will increase to $\sim\!3\times10^7{\rm\,g\,cm^{-3}}$ (for
a cold WD; \citealt{geroh89}).  If the contraction is fast enough to
be (nearly) adiabatic, the central temperature would increase to
$\sim\!1.4\times10^9\,$K (\Fref{rhot}), easily hot enough to ignite.

The timescale for compressional heating is
$\tau_{\rm{}H}=(M_{\rm{}core}/M_{\rm{}disk})\tau_{\rm{}acc}$, where
$M_{\rm{}core}/M_{\rm{}disk}\simeq5$ is the core-to-disk mass ratio, and
$\tau_{\rm{}acc}$ is the disk accretion timescale, which we estimate
using the usual $\alpha$ formalism,
% \citep{shaks73},
\begin{eqnarray}
  \tau_{\rm acc}&=&\frac{M_{\rm disk}}{\dot M} = 
  \alpha^{-1} \left(\frac {r_{\rm disk}}{h}\right)^2 \tau_{\rm dyn}\nonumber\\
  &\simeq& 2{\rm\,h}\;
  \left(\frac{\alpha}{0.01}\right)^{-1} 
  \left(\frac{r_{\rm disk}/h}{4}\right)^2 
  \left(\frac{\Omega}{0.2{\rm\,s^{-1}}}\right)^{-1}.
\end{eqnarray}
Here, $\dot M$ is the accretion rate, and
$r_{\rm{}disk}\simeq0.02\,R_\odot$, $h\simeq0.0055\,R_\odot$, and
$\tau_{\rm{}dyn}=\Omega^{-1}\simeq5{\rm\,s}$ are the disk radius, scale
height, and dynamical time (with numerical values from
\citealt{loreig09}; note that their Table~1 lists outer disk radii;
their Figure~3 shows that most mass is at much smaller radius).

Thus, we find that the remnant is heated on a timescale
$\tau_{\rm{}H}\simeq5\tau_{\rm{}acc}\simeq10\,$h, which is much
shorter than the neutrino cooling and carbon burning timescales even
at the ignition line (where both are $\sim\!3\times10^3\,$yr, see
\Fref{rhot}).  Therefore, neutrino cooling can be ignored, and
compressional heating will continue until the disk is exhausted or the
fusion timescale has become shorter than the accretion timescale.
From \Fref{rhot}, the latter happens when
$\rho\simeq1.6\times10^7{\rm\,g\,cm^{-3}}$ and $T\simeq10^9\,$K, which
is slightly before disk exhaustion (at
$\rho\simeq3\times10^7{\rm\,g\,cm^{-3}}$; see above).  At this point,
a nuclear runaway is inevitable.

\subsection{Complications}
\label{sec:complications}

Above, we argued it is plausible that merger remnants will heat up
sufficiently to ignite carbon, but we made a number of simplifying
assumptions that deserve further study.  First, the ``alpha''
formalism may be inappropriate for estimating the accretion timescale
for a small, massive, and thick disk.  Instead, the relevant timescale
may be the much longer cooling timescale of the envelope
\citep{yoonpr07}.  If transport of angular momentum is the determining
factor, however, our timescale is the correct order-of-magnitude
estimate.  Second, as the accretion rate is highly super-Eddington, a
(strong) wind may form, which may diminish or enhance the compression
depending on its specific angular momentum.  Third, accretion could
heat the envelope significantly.  For cold WDs, rapid accretion leads
to off-center ignition (e.g., \citealt{nomoi84}), but it likely is
less important when the WD core is hot.  Fourth, we assumed the
remnant core rotates roughly uniformly.  This is found in the merger
simulations (\citealt{loreig09}, and references therein), but may
reflect artificial viscosity associated with smooth particle
hydrodynamics.  A strongly differentially rotating remnant would be
less dense and suffer less from compressional heating.

If differential rotation is present, it also leads to additional
effects that we ignored.  It drives a number of processes that tend to
eliminate it \citep{piro08}.  In particular, it will wind up magnetic
fields until their energy is of order the differential rotation
energy, $B\approx(I\Omega\Delta\Omega/R^3)^{1/2}$.  With
$I\Omega\simeq10^{50}{\rm\,g\,cm^2\,s^{-1}}$ and
$R\simeq0.0125\,R_\odot$, the inferred field ranges from
$\sim\!10^9\,$G if differential rotation is driven by the accretion
($\Delta\Omega\simeq\tau_{\rm H}^{-1}\simeq10^{-5}{\rm\,s^{-1}}$) to
$\sim\!10^{11}\,$G if it is due to the merger
($\Delta\Omega\simeq\Omega\simeq10^{-1}{\rm\,s^{-1}}$).  Empirical
evidence for field generation comes from arguably the best candidate
WD merger remnant, RE J0317$-$853.  This WD is massive,
$M\simeq1.35\,M_\odot$,\footnote{In the context of our scenario, the
  existence of this object is puzzling.  It may be the result of an
  unequal mass merger.} spins rapidly, $P=725\,$s, and has a strong,
$B\simeq340\,$MG magnetic field \citep{bars+95}.  Such strong fields,
if they emerge sufficiently fast, would also spin-down the WD through
magnetic dipole emission or through coupling with the accretion disk
and/or a wind (as may have happened for RE~0317$-$853).  The
concomitant loss of rotational support would add to, or might even
dominate, the compressional heating due to accretion.

A separate issue is that to produce SN~Ia, we require detonations.
Spontaneous detonations are very difficult initiate (e.g.,
\citealt{niemw97,seit+09}), although our case is helped by having a
large super-critical region.  Furthermore, at our relatively low
central density, burning is in the distributed rather than flamelet
regime, which may help a possible deflagration transition to a
detonation \citep{niemw97}.

A final complication is the assumed composition.  Merger simulations
so far have used carbon and oxygen only, but $\sim\!1$\% of the mass
will be helium.  If this is burned during the merger, about
$10^{49}{\rm\,erg}$ would be generated, roughly doubling the thermal
energy of the remnant and possibly leading to carbon ignition.

\section{Conclusions}
\label{sec:conclusions}

We have argued that type Ia supernovae result generally from mergers
of CO WDs, even those with sub-Chandrasekhar total mass.  If true, a
number of interesting consequences arise.  First, the merging WDs
should have total masses between double the lowest and highest
possible CO~WD masses, i.e., $1\lesssim M_{\rm
  tot}\lesssim2.4\,M_\odot$ (though ignition during the merger may
cause a break in properties at $\gtrsim\!1.8\,M_\odot$;
\citealt{pakm+10}).  This range could account for
``super-Chandrasekhar'' SN~Ia (e.g., \citealt{howe+06};
\citealt{scal+10}).  As the typical total mass will depend the
population's age, it also explains the empirical age-luminosity
relation.

Second, the rate of mergers---and thus of SN~Ia---may show a break at
$\sim\!1.7\,$Gyr (the lifetime of a $\sim\!1.8\,M_\odot$ star), likely
being more sensitive to the WD formation rate beforehand
\citep{priths08}, and to the merger-time distribution thereafter.
Such a break may have been observed (\citealt{maoz+10}; but see
\citealt{maozsg10}).

Third, ignition will likely take place in rapidly rotating objects,
(possibly) surrounded by small disks.  This may have interesting
consequences for the explosion dynamics \citep{pfanns10}, the
initial shock breakout \citep{pirocw10}, early-time spectra
\citep{mazz+05}, and the supernova remnants.

Finally, we speculate that WDs slowly pushed to the Chandrasekhar
mass---such as should be produced in single-degenerate systems---are
partly responsible for the population of ``atypical'' SN~Ia.  Further
contributions to that population might come from mergers that ignite
during the merger proper, and from unequal mass mergers that ignite
off-center and/or explode only partially.

\acknowledgments This work was triggered by coffee-time astro-ph
discussions at KIAA.  We thank the referee for constructive criticism,
Yanqin Wu for her insistence that rotation was important, Carles
Badenes, Ray Carlberg, Avishay Gal-Yam, Andy Howell, Rubina Kotak,
Kelly Lepo, Yuri Levin, Dany Maoz, and Ken Shen for interesting
discussions, and Enrique Garc\'{\i}a-Berro and Pablo Lor\'en-Aguilar
for providing details on their merger simulations.  We are grateful to
Bill Paxton for making {\sc mesa} publicly available and for setting
it up such that parts can be used separately and ``what if''
experiments are easy to execute.  MHvK and PC thank KIAA and Caltech
Astronomy for their hospitality.  This research made extensive use of
NASA's ADS.

\bibliographystyle{apj}
\bibliography{sniaprog}
\end{document}